\RequirePackage{fix-cm}
\documentclass{svjour3}                     
\smartqed  
\usepackage{graphicx}
\usepackage{color}
\newcommand{\ben}{\begin{enumerate}} \newcommand{\een}{\end{enumerate}}
\newcommand{\beq}{\begin{equation}} \newcommand{\eeq}{\end{equation}}
\newcommand{\beqn}{\begin{equation*}} \newcommand{\eeqn}{\end{equation*}}
\newcommand{\bea}{\begin{eqnarray}} \newcommand{\eea}{\end{eqnarray}}
\newcommand{\ba}{\begin{array}} \newcommand{\ea}{\end{array}}
\newcommand{\beann}{\begin{eqnarray*}} \newcommand{\eeann}{\end{eqnarray*}}
\newcommand{\beasn}{\begin{sneqnarray}} \newcommand{\eeasn}{\end{sneqnarray}}
\newcommand{\bi}{\begin{itemize}} \newcommand{\ei}{\end{itemize}}
\newcommand{\be}{\begin{enumerate}} \newcommand{\ee}{\end{enumerate}}
\begin{document}

\title{Restoration of four-dimensional diffeomorphism covariance in canonical general relativity}
\subtitle{An intrinsic Hamilton-Jacobi approach}

\titlerunning{Diffeomorphism covariant H-J approach to GR}        

\author{Donald Salisbury        \and
        J\"urgen Renn \and Kurt Sundermeyer}


\institute{D. Salisbury \at
              Department of Physics,
Austin College, Sherman, Texas 75090-4440, USA \\
              Tel.: 214 405 6188\\
              \email{dsalisbury@austincollege.edu}             \\
            Max-Planck-Institut f\"ur Wissenschaftsgeschichte,
Boltzmannstrasse 22,
14195 Berlin, Germany \\ 
           \and
           J. Renn and K. Sundermeyer \at
              Max-Planck-Institut f\"ur Wissenschaftsgeschichte,
Boltzmannstrasse 22,
14195 Berlin, Germany
}

\date{Received: date / Accepted: date}

\maketitle

\begin{abstract}
Classical background independence is reflected in Lagrangian general relativity through covariance under the full diffeomorphism group. We show how this independence can be maintained in a Hamilton-Jacobi approach that does not accord special privilege to any geometric structure. Intrinsic spacetime curvature based coordinates grant equal status to all geometric backgrounds. They play an essential role as a starting point for inequivalent semi-classical quantizations. The scheme calls into question Wheeler's geometrodynamical approach and the associated Wheeler-DeWitt equation in which three-metrics are featured geometrical objects. The formalism deals with variables that are manifestly invariant under the full diffeomorphism group. Yet, perhaps paradoxically, the liberty in selecting intrinsic coordinates is precisely as broad as is the original diffeomorphism freedom. We show how various ideas from the past five decades concerning the true degrees of freedom of general relativity can be interpreted in light of this new constrained Hamiltonian description. In particular, we show how the Kucha\v{r} multi-fingered time approach can be understood as a means of introducing full four-dimensional diffeomorphism invariants. Every choice of new phase space variables yields new Einstein-Hamilton-Jacobi constraining relations, and corresponding intrinsic Schr\"odinger equations. We show how to implement this freedom by canonical transformation of the intrinsic Hamiltonian. We also reinterpret and rectify significant work by B. Dittrich on the construction of `Dirac observables'.

\end{abstract}

\section{Introduction}

Few researchers today  believe that an ultimate quantum theory of gravity can or should presuppose any preferred spacetime structure. Geometry must emerge from a background-independent theory. Closely related to this requirement is the foundational property of the general theory of relativity. The classical Einstein equations are covariant under the four-dimensional diffeomorphism group. Yet it is remarkable that even today debate persists on the nature of diffeomorphism covariance of the Hamiltonian formulation of Einstein's theory and the question whether true diffeomorphism invariants (or observables) exist and if so, how does one go about finding them. We need only mention the continuing debate on the ``problem of time" as motivation for this contribution.  We shall briefly show, in as elementary manner as possible, that the efforts of many historical investigators in this field can be understood and reinterpreted in a wider framework. 

Methods for respecting diffeomorphism covariance in a phase space approach to general relativity were first invented in 1930 by L\'eon Rosenfeld \cite{Rosenfeld:1930aa,Rosenfeld:2017ab}, although his significant achievements were largely unknown to the individuals who are now recognized as the creators of constrained Hamiltonian dynamics - Peter Bergmann \cite{Bergmann:1949aa,Bergmann:1949ab}, Paul Dirac \cite{Dirac:1950aa,Dirac:1951aa}, Arnowitt-Deser-Misner (ADM) \cite{Arnowitt:1959aa,Arnowitt:1962aa}, and Karel Kucha\v{r} \cite{Kuchar:1972aa}. Each of these initiated their research with distinctly different motivations. Only in the case of Bergmann did the underlying diffeomorphism covariance play a central role. Dirac's starting point was actually a flat spacetime model in which arbitrary parameterized spacelike hypersurfaces were introduced by hand. ADM dealt with gauge symmetry essentially in the manner that Arnowitt and Deser had learned from their mentor Julian Schwinger - that is to say they attempted from the beginning to work with a preferred spacetime coordinate choice, identified by them as ``intrinsic" coordinates.\footnote{We caution the reader that various notions of ``intrinsic" can be found in the literature (very often also characterized by the term ``internal"). What we have in mind is the idea presented by Komar and Bergmann \cite{Bergmann:1960aa}. It is not to be confused with intrinsic versus extrinsic geometries, fixed in terms of either the 3-geometry or the extrinsic curvature.} Having made a choice they were not concerned with the relation of the resulting theory to models that would result from different choices. Kucha\v{r} continued in the tradition of ADM, showing that at the classical level one could in principle implement a choice of coordinates as a canonical transformation. In this note we intend to show how a suitably improved and reinterpreted version of the Kucha\v{r} program links back to the focus of Bergmann and his group. It turns out that the gauge fixing envisioned by Kucha\v{r} is a construction of true diffeomorphism invariants that never-the-less undergo nontrivial intrinsic temporal evolution.\footnote{See \cite{Salisbury:2021ac} for further explication of this history.} Our result is a suitably modified realization of the partial observable program of Rovelli \cite{Rovelli:2002ab}. Dittrich \cite{Dittrich:2006aa,Dittrich:2007aa}  and Thiemann \cite{Thiemann:2006ac} based their approach to the construction of diffeomorphism invariants on the Kucha\v{r} multi-fingered time program that was in turn based on Dirac's implementation of Hamiltonian constraints. We will show that when properly interpreted from the point of view of the underlying full diffeomorphism symmetry, when sufficient restrictions are placed on the range of permissible partial observables, their constructions can be interpreted as a choice of intrinsic coordinates. We stress that one of our main objectives in this paper is to show how all of these procedures, when suitably implemented and interpreted, can be unified within the framework of the full background-independent diffeomorphism covariance of the original Einstein theory. In addition, we wish to stress that we are not proposing a technical procedure that will necessarily lead to simpler computational techniques. The improvement is one of principle. We are simply elevating diffeomorphism covariance to its original foundational status.

We will begin  in Section 2 by comparing and contrasting a somewhat analogous intrinsic Hamilton-Jacobi approach to a simple toy model, the relativistic free particle. This will serve as both an inspiration and a warning of the inadequacy of this model in addressing general relativity.  One reason we use this example is that it has spawned what we view as an inappropriate terminology in characterizing the construction of observables in general relativity. Several authors refer to these procedures as introducing a ``deparameterization'' of general relativity, based on what they view as an analogous procedure in the artificially parameterized relativistic free particle model. Then in Section 3 we will briefly review a recent construction of diffeomorphism invariants that is an outgrowth of Bergmann's Syracuse school. We explain the necessity of choosing spacetime scalars as intrinsic coordinates and show how invariants result from these choices. We then present in Section 4 a geometrical improvement of the Kuchar program in which a link is established between classical actions corresponding to distinct intrinsic coordinate choices. The first step in the program is accomplished through an appropriate choice of canonical phase space transformations, yielding new fully covariant Hamiltonian field equations. In Section 5 we show that or each field variable choice there corresponds a new Einstein-Hamilton-Jacobi equation. Then for each choice there exists a natural intrinsic coordinate gauge choice. It turns out that the freedom of choice coincides with the original diffeomorphism freedom, and we show  in Section 6 how one can pass from one choice to another with  canonically implemented point transformations. The implications of this freedom are profound. In particular we show  in Section 7 that Bergmann's notion of phase space diffeomorphism equivalence classes, presented in 1961 \cite{Bergmann:1961aa}, is incomplete.  Bergmann's equivalence classes are labeled by $2\times \infty^3$ diffeomorphism invariants, but different choices of intrinsic time and space coordinates leads to distinct evolution equations. The mere identification of orbits swept out through the action of the Hamiltonian is not sufficient either to fix the algebra of observables or determine the evolution of diffeomorphism invariants. We then consider  in Section 8 some semi-classical implications of this program. For every choice of intrinsic canonical variable there exists a corresponding Wheeler-DeWitt equation. But rather than employ this equation it is more efficient to work with an intrinsic Schr\"odinger equation.  In Section 9 we will conclude by illustrating some of these ideas with a simple mini-superspace model.  In Appendix A we illustrate the absence of the parameter evolution parameter in the Hamilton principal function that results from a substitution of general parameterized solutions into the action. Appendix B contains a critique of the Dittrich partial and complete observable program in which the full four-dimensional diffeomorphism symmetry group is lost, in contrast to our treatment. 

\section{The free relativistic particle}

Consider the reparameterization covariant action for a free particle of mass $m$, in units in which $c=1$,
$$
S = -m \int  (-\dot q^2)^{1/2} d\theta = \int L d\theta,
$$
where $q^\mu(\theta)$ represents the spacetime position, and $\dot q^\mu := dq^\mu/d\theta$.
There exists a technique for employing this action to construct a classical and quantum model that establishes a dynamical correlation between observable variables, recognizing that the parameter $\theta$ is itself not observable. The task is to relate this parameter to a measurable physical quantity.
 
The standard point of departure for the Hamilton-Jacobi technique is to consider changes to solutions of the equations of motion that follow from an independent change $\delta_0$ in the configuration variables accompanied by a change in the evolution parameter.   We put some stress on this method since it appears not to be well appreciated that this procedure can be carried out also for singular systems, as in this model. 

In this case of the free particle, the net variation is $\delta q^\mu = \delta_0 q^\mu + \dot q^\mu \delta \theta$.  So the variation at fixed $\theta$ is $\delta_0 q^\mu = \delta q^\mu - \dot q^\mu \delta \theta$. These variations are in fact defined for all $\theta$ in the range of integration of $S$. Consequently we have
$$
\delta S = \int^{\theta_1}_{\theta_0} \left(-\frac{d}{d\theta} \frac{\partial L}{\partial \dot q^\mu } \right) \delta_0 q^\mu + \left( \frac{\partial L}{\partial \dot q^\mu } \left[ \delta q^\mu - \dot q^\mu \delta \theta \right]  + L \delta \theta \right)\left. \right|^{\theta_1}_{\theta_0}.
$$
 The last term on the right comes from the shift in the integration range. When we specialize to the case where no changes are contemplated at $\theta_0$, and assume that at $\theta_1$ we are simply extending the solution to $\theta_1 + \delta \theta$, i.e., $\delta q^\mu(\theta_1) = \dot q^\mu(\theta_1)$ we obtain the increment in the action
\beq
dS = \frac{\partial L}{\partial \dot q^\mu} dq^\mu - \left(\frac{\partial L}{\partial \dot q^\mu} \dot q^\mu - L  \right)  d\theta =: \tilde p_\mu dq^\mu - \tilde H (\dot q) d\theta. \label{dSpart}
\eeq
This is of course a singular system; the momenta $\tilde p_\mu (\dot q):=\frac{\partial L}{\partial \dot q^\mu} =  m\dot q_\mu (-\dot q^2)^{-1/2}$ are not independent. In fact, since the Lagrangian is homogeneous of degree one in the velocities, $\tilde H := \frac{\partial L}{\partial \dot q^\mu} \dot q^\mu - L \equiv 0$. The constraint takes the form $H  =  \frac{1}{2m}(\tilde p^2 + m^2) \equiv 0$. In Appendix A we show how one can employ (\ref{dSpart}) to construct a complete set of Hamilton principal functions $S(q^\mu,\theta; \bar p)$ that satisfy the Hamilton-Jacobi equation
\beq
\frac{\partial S}{\partial \theta} + \tilde H \left(q, \frac{\partial S}{\partial q};\theta  \right) = 0. \label{HJpart}
\eeq
And we show that one can in the usual manner employ this principal function to obtain the general solution of the parameterized free particle trajectories.

But rather than regain the parameterized trajectories we will inquire as to how one can  gain information from $dS$ on the measurable physical evolution of the single particle system - recognizing that the parameter $\theta$ is ostensibly not a physically measurable quantity. In this case the answer is clear. One could simply choose the reparametrization scalar $q^0$ as the evolution time. In doing so a relation is established between the in principle measureable spatial position of the particle and the measurable Minkowski time $q^0$. This is a choice of intrinsic time - intrinsic in the sense that the evolution parameter itself is measurable.  As we shall see, it is essential that our choice transform as a scalar under the action of the reparameterization group. There is in general a two step procedure  for making a choice of intrinsic coordinates. The first step is to make a canonical change of variables. The second step is to choose one of these variables as the evolution time.
 
In this particular case the first step is already accomplished since $q^0$ is already a configuration space variable. But in order to stress the fact that the isolation of the $q^0(\theta)$ does not automatically imply that one has made an intrinsic coordinate choice we note that this variable does undergo variations under the reparameterizations of the form $\theta' = \theta + \delta \theta = \theta - (-\dot q^2)^{-1/2} \xi(\theta)$.\footnote{The $(-\dot q^2)^{-1/2} $ dependence of the reparameterizations results from the requirement that variations in $q^\mu$ be projectable under the Legendre map to phase space. The generator is the vanishing charge that results from Noether's second theorem. See \cite{Salisbury:2021ad} for details.} The phase space generator of these variations is $G(\xi) = \xi H$. Since the following will play a role in the comparisons with the literature, we display explicitly the action of the reparameterization generator on the variable $q^0$. Under this infinitesimal reparameterization the corresponding variation of $q^0$ is
$$
\bar \delta q^0 := q'^0 - q^0 = \left\{q^0, H \xi  \right\} = \xi p^0 = \dot q^0 (- \dot q^2)^{-1/2} \xi = - \dot q^0 \delta \theta,
$$  
thus recovering precisely the required transformation property of a scalar.

Before turning to the second step, note that since we have not yet made a parameter choice we still have a phase space constraint $H=0$. Then it is natural to ask what would be the consequence of representing the momenta in the constraint as $p_\mu = \frac{\partial S}{\partial q^\mu}$ and interpreting the constraint as a differential equation to be satisfied by $S$. In other words, look for solutions of
\beq
\eta^{\mu \nu} \frac{\partial S}{\partial q^\mu} \frac{\partial S}{\partial q^\nu} + m^2 = 0. \label{parthj}
\eeq
It is significant that solutions of this equation do not give directly solutions $q^\mu(\theta)$ of the reparameterization covariant Euler-Lagrange equations. The parameter $\theta$ does not even appear in this equation! Additional information is required.
Given a solution $S$, one can set
\beq
p_\mu = \frac{\partial S}{\partial q^\mu},
\eeq
Then one must appeal to the Hamiltonian equation $\dot q^\mu = \lambda(\theta) p^\mu$ where one picks the function $\lambda$, obtaining the first order differential equation $\dot q^\mu = \frac{\partial S}{\partial q^\mu}$ which can then be integrated. The point is that only when this function has been selected has one made a choice of gauge. In other words, the ``Hamilton-Jacobi equation" (\ref{parthj}) continues within this formalism to be a constraint, and we have simply managed to solve the constraint. This is the reason that we have enclosed the expression in quotation marks. It is not a true Hamilton-Jacobi equation.

Of course, what really motivates interest in this example is the means that is available to find $q^a$ as a function of $q^0$. In other words, one wants to make an explicit intrinsic parameter choice. And there is a natural way of doing this using the ``Hamilton-Jacobi equation". It does give us directly $S$ as a function of $q^a$, $q^0$ and of three independent constants $\bar p^a$. And solutions for $q^a$ as a function of $q^0$ can be obtained in the usual manner in Hamilton-Jacobi theory by taking derivatives $\frac{\partial S}{\partial \alpha^a}$. Thus the ``Hamilton-Jacobi equation" brings with it a natural choice of intrinsic parameter - due to the fact that this natural choice is one of the configuration variables! As we shall see, this is not true in general relativity, and one of the objectives of this paper is to overcome this difficulty.

It is also possible to make the intrinsic parameter choice directly in the action increment $dS$. We simply interpret $q^0$ as the evolution parameter and the momenta as phase space variables - subject of course to the constraint. Thus we can solve for the intrinsic Hamiltonian $H_i$, setting $p_0 = - \left(p^a p_a + m^2  \right)^{1/2}:= -H_i$, so that the increment in the action in intrinsic coordinates is
$$
dS = - H_i dq^0 + p_a dq^a.
$$
From this expression we deduce the true (intrinsic) Hamilton-Jacobi equation
$$
\frac{\partial S}{\partial q^0} +  \left(\frac{\partial S}{\partial q^a} \frac{\partial S}{\partial q_a}+ m^2  \right)^{1/2} = 0.
$$
It's complete solution is
$$
S = - \left(\vec \alpha^2 + m^2 \right)^{1/2} q^0 + \alpha_a q^a - f(\vec \alpha),
$$
from which one obtains the general classical solution by setting $0 = \frac{\partial S}{\partial \alpha_a}$.
It follows as a consequence of the intrinsic Hamilton-Jacobi equation that the quantum wave function $\Psi(q^0, \vec q) = \int d^3\! \vec \alpha F(\vec \alpha) e^{- S/\hbar}$ satisfies the Schr\"odinger equation
\beq
H_i\left(- i \hbar \frac{\partial}{\partial q^a} \right) \Psi = i \hbar \frac{\partial}{\partial q^0} \Psi. \label{freeSch}
\eeq
Furthermore, again as a consequence of the intrinsic Hamilton-Jacobi equation, an appropriately peaked superposition over $\vec \alpha$ will deliver a correct semi-classical wave packet.

Although it is obvious that the intrinsic dynamics does not depend on the parameter $\theta'$, and is thus invariant under reparameterizations, it is instructive to see how this intrinsic choice yields variables expressed in terms of the $q^\mu(\theta')$ in an arbitrary parameterization but which are invariant under reparameterizations.  They are the observables. The observables associated with $q^a$ are (See \cite{Pons:2009ab})
\beq
{\cal O}_{q^a} = q^a(\theta') - \frac{p^a}{p^0} q^0(\theta') + \frac{p^a}{p^0} \theta, \label{invq}
\eeq
while ${\cal O}_{p^\mu} = p^\mu$. 

The coefficients of each power of $\theta$ are invariant under the active transformations generated by $G(\xi)$. The invariant variables ${\cal O}_{q^a}$ are in fact the $q^a$ in (\ref{freeSch}), while the $\theta$ that appears on the right in (\ref{invq}) is $q^0$.   This means that when one goes to the quantum theory and considers wave functions $\psi({\cal O}_{q^a},\theta)$, these wave functions will satisfy $\hat H \psi({\cal O}_{q^a},\theta) = 0$, where $ \hat H $ is the operator $\frac{1}{2m}\left(\hat p^2 + m^2\right)$ with $\hat p_\mu = - i \hbar \frac{\partial }{\partial q^\mu(\theta')}$. 

Finally, we note that the  observables satisfy a modified Poisson bracket algebra. It was shown in general in \cite{Pons:2009aa} that this algebra is the Dirac bracket algebra constructed using the gauge fixing constraint. In this case the modified bracket is
\beq
\left\{ f,g  \right\}^* = \left\{f, g  \right\} + \left\{f, \frac{1}{2}\left(p^2 + m^2  \right)  \right\}\frac{1}{p^0}\left\{t - q^0, g  \right\},
\eeq
and it is indeed the case that
\beq
\left\{ q^a,p^0  \right\}^* = \left\{ {\cal O}_{q^a},{\cal O}_{p^0}  \right\} = \frac{p^a}{p^0}. \label{qa}
\eeq
Canonical quantization would then require a realization of this algebra. This can be achieved in a momentum representation with $p^0 = \left( \vec p^2 + m^2 \right)^{1/2}$. Interestingly, the operator $\hat q^a$ is in this representation none other than the Newton-Wigner position operator \cite{Newton:1949aa}.

Analogues for all of these constructions can be undertaken in general relativity. But we stress again two significant differences in the treatment of the relativistic particle as compared to general relativity: 1)  $q^0$ is already a reparameterization scalar. Finding such scalars in general relativity is a non-trivial task that the particle model offers no clues in completing  2) It was not adequately realized that just as was clear in the particle model, after taking the first step in general relativity, one still has a theory that is  covariant under the full diffeomorphism group.

\section{Review of Hamiltonian constrained dynamics, observables, and gauge fixing} \label{constrained}

We briefly review here the nature of the diffeomorphim-induced canonical transformation group of general relativity, and the manner in which it can be employed to construct diffeomorphism invariants. \footnote{See \cite{Pons:2009aa}, and literature cited therein for complete details}. The  phase-space realizable infinitesimal general coordinate transformations are
\beq
x'^\mu = x^\mu - n^\mu \xi^0(x) - \delta^\mu_a \xi^a(x),
\eeq
where $n^\mu = \left(N^{-1}, - N^{-1} N^a \right)$ is the normal to the $t = $ constant spacelike hypersurface.
Thus there is an explicit dependence on the metric lapse $N$ and on the metric shift $N^a$, and due to the appearance of the 3-metric $g_{ab}$ in the commutator algebra the descriptors $\xi$ must also depend on the 3-metric. Variations of the canonical variables (including the lapse and shift) under these transformations are generated by\footnote{See also \cite{Salisbury:2021ad} for a simplified derivation of this generator.} \footnote{When there  are additional gauge fields there are additional terms}  \footnote{We employ B. DeWitt's compact notation in which repeated indices also represent an integration}
\beq
G_{{\xi }}(x^0) =  \int d^3 x\,\left(P_\mu \dot\xi^\mu + ( {\cal H}_\mu \label{Gxi}
+N^{\rho'} C^{\nu''}_{\mu\rho'} P_{\nu''} )\xi^\mu  \right),
\eeq
 where the $P_\mu$ are the vanishing momenta conjugate to $N^\mu$, the ${\cal H}_\mu=0$ are the secondary constraints and the $C^\nu_{\mu\rho}$ are the structure coefficients of the Dirac algebra. This generator is to be distinguished from the generator of time evolution,
\beq
H = \int d^3\!x \, \left(N^\mu {\cal H}_\mu + \lambda^\mu P_\mu \right). \label{hgrav}
\eeq
The $\lambda^\mu$ are almost arbitrary spacetime functions, the only restriction being that $\lambda^0 > 0$.

The generator $G_{{\xi }}(x^0)$ is deployed to construct diffeomorphism invariants by choosing coordinate conditions.  We fix coordinates by employing physical landmarks in spacetime. Specifically, we correlate spacetime events with specific values of the spacetime curvature. Such a coordination is called ``intrinsic" since it can be used to establish a correlation between a subset of field variables and the remaining dynamical variables.\footnote{This is in the spirit of Einstein's point coincidence argument.} We then actively map within a given arbitrary chart to events that satisfy  the chosen intrinsic coordinate conditions. So as not to cause unnecessary confusion in the following we will refer to the coordinates in the fixed chart as ``coordinate parameters". These will be distinguished from ``intrinsic coordinates". The invariants that we obtain through this procedure are invariant under arbitrary active maps of the original coordinate parameter chart. 

We give a simple argument why the intrinsic coordinate functionals must be spacetime scalars - as was recognized by Bergmann, Bryce DeWitt, and ADM. We suppose that the parameters $x^\mu$ have been fixed by the condition that $x^\mu = X^\mu(g(x), p(x))$. We investigate how this relation transforms under a change of parameters $x'^\mu = f^\mu(x)$. Then we must have
\beq
x'^\mu = f^\mu(x) = f^\mu\left(X(g(x), p(x)\right) = f^\mu\left(X(g'(x'), p'(x'))\right), \label{xprime}
\eeq
where on the right we first perform the coordinate parameter transformations before mapping to intrinsic coordinates. Since the result must not depend on the order in which these operations are performed we
deduce from (\ref{xprime}) that since $f$ is arbitrary,
\beq
X^\mu(g(x), p(x)) = X^\mu(g'(x'), p'(x')).
\eeq
In other words, the $X^\mu$ must be spacetime scalar functions.\footnote{Kucha\v{r}  \cite{Kuchar:1972aa} required only that the $X^\mu$ be scalars under spatial diffeomorphisms. Dittrich and Tambornino \cite{Dittrich:2007ab,Dittrich:2007ac} discuss technical advantages and disadvantages in employing spacetime scalars as partial variables. } It has been shown that if the scalar condition is satisfied then no physical solutions are eliminated, and if is not satisfied then the fixation of coordinates is not unique.\cite{Pons:2010aa} We of course have in mind generic spacetimes with no Killing fields and we recognize that generally one must patch together intrinsic coordinate charts.

In general relativity with material sources we have at our disposal at least fourteen scalars that can be constructed from the Riemann curvature tensor \cite{Zakhary:1997aa}. They generally involve quadratic or cubic powers of this tensor. 

It was shown in \cite{Pons:2009aa} how to construct the observables that correspond to given intrinsic coordinate choices $x^\mu = X^\mu[g_{ab}, p^{cd}]$. The outcome is that there corresponds to every phase space variable (including the lapse and shift) a series expansion in powers of the intrinsic coordinates, the coefficients of which are invariant functionals of the metric.\footnote{Dittrich derived analogous infinite series, with invariant coefficients, referring only to the three-metric and conjugate momenta \cite{Dittrich:2006aa}. It was shown in \cite{Pons:2009ab} how to construct analogous series through active transformations also for the lapse and shift. } These coefficients are invariant in the conventional sense that variations of the metric under changes of the coordinate parameters does not change their values.

One outcome of this construction that has as yet not received attention in the literature is that once one has found an acceptable set of spacetime scalar functionals $X^\mu(g(x), p(x))$ then one is in position to solve in any coordinate parameter chart for the coordinate parameters in this chart as functionals of $g_{ab}$ and $p^{cd}$. The solution proceeds as follows: Let us suppose that we are in possession of a particular solution of Einstein's equations expressed in coordinate parameters $x^\mu$. To solve for the $x^\mu$ as functionals of $g_{ab}$ and $p^{cd}$ we most simply find the coordinate transformation that transforms from the intrinsic coordinates $X^\mu[g_{ab}(x), p^{cd}(x)]$ to the coordinates $x^\mu$, i.e. $x^\mu = f^\mu \left( X[g_{ab}(x), p^{cd}(x)] \right)$. This expression constitutes the solution we are seeking. We will illustrate this procedure below for a cosmological example.

We now show how the construction of diffeomorphism invariants is related to a completion of Kucha\v{r}'s scheme for introducing what he calls internal coordinates and a corresponding reduced Hamiltonian.

\section{Intrinsic Hamiltonians}

We will ultimately describe a formalism in which the gravitational action will be a functional of $2\times \infty^3$ independent metric variables which are functions of intrinsic coordinates, and a corresponding number of constants. The task of constructing such an action is complicated by the fact that the curvature-based spacetime scalars we shall use depend on time derivatives of the metric variables. Because of this the introduction of these scalars directly into the gravitational Lagrangian would result in the appearance of second order time derivatives. It may be the case that no intrinsic Lagrangian exists, as is true for the simple cosmological example we consider below.

Actually, although he did not address this particular issue, Karel Kucha\v{r}'s pioneering work \cite{Kuchar:1972aa,Kuchar:1973aa} suggests a way to proceed - via the appropriate canonical phase space transformation. We will apply the procedure to the vacuum case. The generalization to the non-vacuum case is straightforward. It is noteworthy that Kucha\v{r} viewed his procedure as the selection of a family of dynamical-field dependent embeddings of spatial hypersurfaces into spacetime. In introducing the notion of ``bubble time" or "multi-fingered time", he explicitly rejected the idea that the full four-dimensional diffeomorphism group could be realized in the Hamiltonian formulation of the general theory of relativity. We shall show that a reinterpretation of his method does bring the full group back into play.

We begin by writing the increment in the action as
\beq
dS_{GR} = \int d^3\!x \left( \tilde p^{ab} dg_{ab} + \tilde P_\mu dN^\mu\right) -dt \int d^3\!x \, \left(N^\mu {\cal H}_\mu + \lambda^\mu P_\mu \right), \label{action}
\eeq
where the tilde signifies that the momenta are to be conceived as configuration-velocity functions.  As in the particle example we consider net variations to new solutions that result from $\delta_0$ variations of the fields $g_{\mu \nu}$ and variations in this case of the four coordinate parameters.\footnote{The field-theoretic Hamilton-Jacobi approach (for electromagnetism) was first formulated in this manner by Paul Weiss \cite{Weiss:1936aa}. The application to vacuum general relativity proceeds as follows. Let $\delta g_{\mu \nu}$ represent the net variation. Then we must subtract from it the variation that results from $\delta x^\mu$ to obtain the variation at fixed $x^\mu$. The resulting variation that takes into account the shift $\delta x^0$ is 
$$
\delta S_{GR} =  \int d^3\!x \left[  p^{ab}\delta_0 g_{ab} +P_\mu \delta_0 N^\mu -\left(p^{ab}\dot g_{ab} - {\cal L}_{GR} + \pi_\mu  \dot N^\mu   \right) \delta x^0 \right. $$ . 
}
This is the standard Hamilton-Jacobi procedure, but now applied to a singular dynamical system. Of course the $\tilde P_\mu$  are vanishing primary constraints. And we must satisfy the secondary constraints  ${\cal H}_\mu = 0$.  

Our first task in extending Kuchar's work is to consider the phase space formulation of the action increment  (\ref{action}), and then to implement a canonical phase space transformation in which we introduce the $X^\mu[g_{ab}, p^{cd}]$ as canonical variables. Contrary to Kuchar we do not permit the $X^\mu$ to depend explicitly on the coordinate parameters  precisely because we insist that we are always choosing a coordinate system based on measurable curvatures. We can then rewrite the non-vanishing contributions to the action increment in terms of the intrinsic coordinates and their conjugate momenta. Thus we seek  canonical transformations such that
\bea
dS_{GR} &=& \int d^3\!x \left( p^{ab} dg_{ab} + P_\mu dN^\mu\right) - dx^0 \int d^3\!x \left(N^\mu {\cal H}_\mu \left[g_{ab},  p^{cd} \right]+ \lambda^\mu P_\mu \right)\nonumber \\
&=&  \int d^3\!x \,\left( \pi_\mu dX^\mu +   p^{A} dg_{A} +\frac{\delta G}{\delta g_{ab}} dg_{ab} +\frac{\delta G}{\delta g_A} dg_A+\frac{\delta G}{\delta X^\mu} dX^\mu  +  P_\mu dN^\mu \right) \nonumber \\
&-&dx^0 \int d^3\!x \left(N^\mu {\cal H'}_\mu\left[g_A, p^B, X^\mu, \pi_\mu\right]+ \lambda^\mu P_\mu \right).
\eea
(The necessity of a transformation of the action with a corresponding generating function $G$ is not mentioned explicitly in Kuchar's work.)
 We assume that the $X^\mu$ commute with each other and with the new phase space variables $g_A, p^B, A, B = 1,2 $ as well. The $\pi_\mu$ are canonically conjugate to the $X^\mu$ while the $p^A$ are canonically conjugate to the $g_A$, and they are obtained through the generating functional such that,
\beq
p^{ab} = \frac{\delta G}{\delta g_{ab}},
\eeq
\beq
\pi_\mu =- \frac{\delta G}{\delta X^\mu},
\eeq
and
\beq
p^A = - \frac{\delta G}{\delta g_A}.
\eeq
Thus we have the canonical change of variables $(g_{ab}, p^{cd}) \rightarrow (X^\mu, g_A, \pi_\nu, p^B)$. It must be emphasized that the transformed action is still fully covariant under the action of the diffeomorphism-induced symmetry group whose generators (\ref{Gxi}) are now expressed in terms of the new canonical phase space variables, with corresponding changes in the structure coefficients of the Dirac algebra.

The canonical transformations we have just undertaken are the analogue of the first step in the free particle model. This step was trivial in the particle model with the gauge choice $\theta = q^0$ since $q^0$ was already a configuration variable. It was less trivial with the proper time gauge choice.
The advantage that we have gained through this change of variables is that we can now easily isolate in the non-vanishing contribution to the action increment those variables that will serve as our intrinsic coordinates. We will later,  in our second step, let $x^\mu = X^\mu$. Finally, in order to respect the constraints,   we solve the  constraints $0 = {\cal H}'_\mu \left(g_A,x^\mu,p^B,\pi_\nu  \right) $ for $\pi_\nu \left( g_A, p^B, x^\mu \right)$ whereby the   $\pi_\nu$ become explicit functionals of $g_A, p^B$ and $x^\mu$. Then since there is no incremental change in the $X^a$ in this gauge, the non-vanishing contribution to the action becomes the intrinsic canonical one-form
\beq
\theta_{ i} = \int d^3\!x\, \left( p^{A} dg_{A} + \pi_0\left[g_A, p^B, x^\mu\right] dx^0 \right). \label{thetaintrin}
\eeq
Thus we deduce that 
\beq
H_{i} := - \pi_0 \left[g_A, p^B, x^\mu\right], \label{intrinsicham}
\eeq
is the intrinsic Hamiltonian\footnote{ in the literature also known as the ``reduced", ``true"  or ``physical'', Hamiltonian; we prefer to call it the intrinsic Hamiltonian in order to stress that it is built with the genuine  measurable space-time scalars }, while the $\pi_a \left[g_A, p^B, x^\mu\right]$ are canonical generators of spatial displacements.

It is important to notice that even though the lapse and shift have disappeared from this formalism, they are in fact fixed through the condition that the intrinsic coordinate choice be preserved under time evolution. They become fixed functionals of $g_A, p^B$ and $x^\mu$.

\section{Hamilton-Jacobi formalism}

After making the change of canonical variables, but before fixing a gauge by choosing intrinsic coordinates, we are in position to rewrite  what has been called the Einstein-Hamilton-Jacobi equation using these new variables. In the transformed Hamiltonian constraint $ {\cal H'}_\mu\left[g_A, p^B, X^\mu, \pi_\mu\right] = 0$, analogously to Peres' original proposal \cite{Peres:1962aa} , we simply write
$$
{\cal H'}_\mu\left[g_A, \frac{\delta S}{\delta g_B}, X^\mu, \frac{\delta S}{\delta X^\nu}\right] = 0.
$$
As was demonstrated by Gerlach using conventional metric variables \cite{Gerlach:1969aa}, from solutions of this equation one can construct solutions of Einstein's equations. However, specific choices for the lapse and shift variables must still be made.\footnote{See for example \cite{DeWitt:1970aa} }. On the other hand there are natural choices corresponding to trivial functions; these are namely lapse equal to one and vanishing shift in the conventional case. The trivial choices using the new phase space variables actually yield the intrinsic dynamics. We claim, in other words, that the Einstein-Hamilton-Jacobi constraint equation in the new variables is equivalent to the true Hamilton-Jacobi equation obtained using the intrinsic Hamiltonian  (\ref{intrinsicham}). 
We call this the true Hamilton-Jacobi equation since it is obtained in the usual manner by seeking the canonical transformation that produces a vanishing Hamiltonian. Calling the generator of this transformation $\bar S$, the condition that the resulting Hamiltonian vanishes is
\beq
H_{i}\left[g_A, p^B, x^\mu\right] + \frac{\partial \bar S}{\partial t} = 0 = H_{i}\left[g_A, \frac{\delta \bar S}{\delta g_B}, x^\mu\right] + \frac{\partial \bar S}{\partial t}.
\eeq
We will demonstrate the equivalence of this equation with the transformed Einstein-Hamilton-Jacobi equation below for a  simple cosmological model.

\section{Intrinsic coordinate transformations}

As we have stressed above, having made a particular choice of intrinsic coordinates we now have the liberty to undertake an arbitrary finite transformation to new intrinsic coordinates $X'^\mu = f^\mu(X)$. This is merely a point transformation, and can therefore be realized in phase space as a canonical transformation. We now derive the corresponding new intrinsic Hamiltonian. We require that 
\beq
 \int d^3\!x \,  X^\mu d\pi_\mu = - \int d^3\!x  \pi'_\mu dX'^\mu +  \int d^3\!x d F \left[X', \pi \right].
\eeq
It follows that
\beq
X^\mu = \frac{\partial F}{\partial \pi_\mu} = f^{-1}{}^\mu(X'), 
\eeq
and
\beq
\pi'_\mu = \frac{\delta F}{\delta X'^\mu},
\eeq
and therefore
\beq
F = \int d^3\!x \, f^{-1}{}^\mu(X') \pi_\mu, 
\eeq
and the transformed intrinsic Hamiltonian is \footnote{Kucha\v{r} did not address this issue. We have here the rule for canonically transforming his reduced Hamiltonian.}
\beq
H'_{i} = -\pi'_0 = -\frac{\delta F}{\delta X'^0}.
\eeq

\section{Diffeomorphism equivalence classes and intrinsic dynamics}

It has long been an widespread belief that diffeomorphism equivalence classes of solutions of Einstein's equations can be described exclusively in terms of the phase space variables $\left(g_{ab}, p^{cd}  \right)$. This is incorrect. The error stems from a common conflation of two distinct notions in constrained Hamiltonian dynamics for reparameterization covariant systems, namely time evolution on the one hand, and diffeomorphism symmetry on the other. In the Hamiltonian formalism time evolution, i.e., global rigid translation in time is not a canonically realizable symmetry. More precisely, there does not exist a member of the diffeomorphism-induced symmetry group that affects a global rigid time translation on \underline{every} solution of Einstein's equations. Rather, a group element will perform this feat only on one particular solution. On the other hand the diffeomorphism-induced symmetry group transforms solutions of Einstein's equations into new solutions. The difference is particularly evident when discussing  the dynamical evolution of the intrinsic variables. In this case the analysis of section \ref{constrained} shows that corresponding to every acceptable spacetime  scalar  intrinsic coordinate choice the variables $g_A$ and $p^B$ can be displayed as manifestly invariant functionals under the action of the full diffeomorphism-induced symmetry group. These manifestly invariant functionals are constructed by performing finite symmetry transformation to that location on symmetry orbits (for each value of the coordinate parameters $x^\mu$) at which the intrinsic coordinate conditions are satisfied. On the other hand one obviously has a nontrivial time evolution.  The $2 \times \infty^3$ values of $p^A(\vec x)$ at a fixed intrinsic coordinate time fix an equivalence class. These are the analogues of the initial momenta $\bar p^a$ in the free relativistic particle model. And as in that model, the evolution in intrinsic time depends on the choice of intrinsic coordinates. It is not sufficient as is commonly maintained to claim that phase space equivalence classes are simply orbits under the action of the diffeomorphism symmetry group in $\left(g_{ab}, p^{cd}  \right)$ phase space. There is an accompanying non-trivial evolution in intrinsic time. 

A remark concerning conventional geometrodynamics is appropriate here. In this program one in effect grants three-dimensional space a preferred geometrical status, and one similarly boosts this status by emphasizing the role of the three-dimensional spatial diffeomorphism group. But as we have seen, it is not possible to assign intrinsic spatial spacetime curvature-based coordinates without making reference to the temporal continuation of the three-metric off the spatial hypersurface - since all curvature scalars depend on the three-momentum.

\section{Semi-classical quantization}

The standard approach to the semi-classical quantization of gravity is via the Wheeler-DeWitt equation where one replaces the 3-momentum variables in the conventional ${\cal H}_0$ constraint by the operator $-i \hbar \frac{\delta}{\delta g_{ab}}$. But we now have at our disposal an infinite multitude of ${\cal H}_0$ constraints, one for each choice of the $X^\mu$.  For each selection there is a corresponding Wheeler-DeWitt equation, each of which will yield its own natural choice for intrinsic temporal and spatial coordinates. However,  as we shall argue in the cosmological example below in referring to the intrinsic Hamilton-Jacobi equation, the physical content of these Wheeler-DeWitt equations must already be contained within an intrinsic Schr\"odinger equation obtained by making the canonical operator substitutions in the intrinsic Hamiltonian. Thus we have for each choice of the $X^\mu$ an intrinsic Sch\"odinger equation
\beq
H_{i}\left[g_A, - i \hbar \frac{\delta}{\delta g_B}; x^\mu  \right] \Psi\left[g_A, x^\nu \right] = i \hbar \frac{\partial }{\partial t} \Psi\left[g_A, x^\nu \right].
\eeq
We are assured that we can construct the correct semiclassical limit, for each of the choices for $X^\mu$, from solutions of the form $\Psi = \sigma e^{i S/\hbar}$ where $S$ is a complete solution of the intrinsic Hamilton-Jacobi equation.

It is remarkable that the independent $g_A$ and $p^B$ are canonically conjugate and yet they are to be understood as diffeomorphism-induced invariants. However, when one undertakes the construction of invariants in the manner described above, the resulting invariants will satisfy the Dirac bracket algebra.  They will generally not be canonically conjugate.  It is the case however that non-canonical transformations can be undertaken so that this subset does satisfy the canonical Poisson bracket algebra. 

Following up on our critique in the previous section of the  conventional geometrodynamical program, one cannot anticipate that appropriate choices of spacetime curvature based intrinsic coordinates can emerge from the  usual Einstein-Hamilton-Jacobi equation, and its associated Wheeler-DeWitt equation. The reason is that in this standard approach the action is viewed as a functional of the three-metric. And as we have seen, neither spatial nor temporal spacetime scalars can be constructed employing the three-metric alone.

\section{A cosmological example}

We illustrate these ideas with a simple cosmological example, an isotropically expanding universe with vanishing curvature, vanishing cosmological constant, and a scalar source field. The line element is
\beq
ds^2 = -N(t)^2  dt^2 + a(t)^2  (dx^2 +  dy^2 +  dz^2),
\eeq
with Lagrangian
\beq
L = \frac{1}{2 N}\left(-\frac{6}{\kappa } a \dot a^2 + a^3 \dot \Phi^2  \right).
\eeq
The corresponding canonical Hamiltonian is
\beq
H_c  = N \left( -\kappa \frac{ p_a^2}{12 a} + \frac{ p_\Phi^2}{ 2 a^3}\right),
\eeq
where $\kappa:= 8 \pi G$, and we have the  weakly vanishing secondary constraint
\beq
H= -\kappa \frac{ p_a^2}{12 a} + \frac{ p_\Phi^2}{2 a^3}  = 0. \label{hcosmo}
\eeq
We find that the quadratic Riemann scalar \footnote{$R^1$ is not to be confused with $R_{\alpha \beta \gamma \delta}R^{\alpha \beta \gamma \delta}$.}
\beq
R^{1} := R_{\alpha \alpha' \beta \beta' } g^{\beta \beta'
\gamma \gamma'}
R_{\gamma \gamma' \delta \delta' } g^{\delta \delta' \alpha \alpha'}, \label{r1}
\eeq
where
$
g^{\beta \beta' \gamma \gamma'} := 2 g^{\beta [\gamma} g^{\gamma'] \beta'},
$
simplifies to a power of $a^2 p_a^{-1}$ for this highly symmetric solution. \footnote{Bergmann and Komar's procedure \cite{Bergmann:1960aa} shows that in general some of the curvature scalars's can be expressed in terms of the 3-metric and conjugate momenta. In particular when expressed in terms of these phase space variables there is no explicit dependence on the lapse and shift}  We therefore take $T = \frac{1}{3}\kappa a^{-2} p_a$, \footnote{With this choice $T$ will range from $-\infty$ to 0}, as one of our new phase space variables. Note that since $\dot a = - \frac{\kappa N}{6a}  p_a$, this variable is actually minus twice the Hubble parameter 
$ N^{-1} a^{-1} \dot a$. The choice we make here is actually proportional to the extrinsic curvature scalar, and is also known as the York time \cite{York:1972aa}. \footnote{See \cite{Roser:2014aa} for a proposed use of York time for this cosmological model.}

The variable conjugate to $T$ is $p_T =- \frac{1}{\kappa} a^3$ so that the corresponding generator in
\beq
p_a da = p_T dT + \frac{\partial G}{\partial a} da + \frac{\partial G}{\partial T }dT
\eeq
 is
\beq
G = \kappa^{-1} a^3 T.
\eeq
The inverse canonical transformation is $a =\kappa^{1/3}\left( -p_T  \right)^{1/3}$ and 
$$p_a = 3 \kappa^{-1/3} \left( - p_T  \right)^{2/3}T.$$ 
In terms of the new canonical variables the constraint (\ref{hcosmo}) becomes
\beq
H' = \frac{3}{4}   p_T T^2 - \frac{p_\phi^2}{2 \kappa p_T} = 0. \label{hprime}
\eeq
The generator of the Legendre-projectable reparameterizations $ t' = t - N(t)^{-1} \xi(t) $ in terms of these new variables is
\beq
G'_\xi (t) = \xi(t) H'(t)
\eeq
The equations of motion are
\beq
\dot T \approx  \frac{3}{4} N T^2, \,\,\, \dot p_T = -  N \frac{3}{4} p_T T, \label{dotT}
\eeq
and
\beq
\dot \phi = -N \frac{1}{2}\kappa^{-1} \frac{p_\phi}{p_T} \approx -  \kappa^{-1/2} N\frac{1}{2} \left(\frac{3}{2}  \right)^{3/2}\left(-T  \right), \,\,\, \dot p_\phi = 0.
\eeq

Of course, having managed to write the action increment in the form
\beq
dS = p_T dT + p_\phi d\phi, 
\eeq
we can immediately implement our choice of intrinsic coordinate by simply letting $T = t$ in this expression and also solving the constraint for the momentum, obtaining
\beq
p_T = - H_{i} = \frac{1}{\left(3 \kappa/2\right)^{1/2}} \frac{p_\phi}{\left(-t \right)}. 
\eeq
 Substituting into the action we obtain the intrinsic action
\beq
S_{i} = - H_{i} dt + p_\phi d\phi.
\eeq
This leads to the simple equation of motion
\beq
\dot \phi = - \frac{1}{\left(3 \kappa/2\right)^{1/2} (-t)},
\eeq
with solutions
\beq
\phi = \phi_0 + \frac{1}{\left(3 \kappa/2\right)^{1/2}} \ln \left(- t) \right), \label{solution}
\eeq
as $t$ ranges from $-\infty$ to $0$.

Finally, to find the lapse we must return to the original Hamiltonian equations, we set $\dot T = 1$ in (\ref{dotT}), obtaining $N = \frac{2}{3 t^2}$.

Note also that the Einstein-Hamilton-Jacobi equation from the constraint (\ref{hprime}) is
\beq
 \frac{3}{2}  \left(\frac{\partial S}{\partial T}\right)^2 T^2 - \frac{1}{\kappa} \left(\frac{\partial S}{\partial \phi}\right)^2 = 0. \label{ehjeq}
\eeq
The intrinsic Hamilton-Jacobi equation is therefore the square root of the Einstein-Hamilton-Jacobi equation (\ref{ehjeq}),
\beq
 \frac{\partial S_i}{\partial t}  + \frac{1}{\left(3 \kappa/2\right)^{1/2}} \frac{1}{\left( t + 1 \right)}\frac{\partial S_i}{\partial \phi} = 0.
\eeq
The complete solution is
\beq
 S_i(\phi,t;\alpha) = e^{-\alpha\phi_0 - \frac{\alpha}{\left(3 \kappa/2\right)^{1/2}} \ln (-t) + \alpha \phi}
\eeq
One obtains the general classical solution for $\phi$ in the usual manner by setting $0 = \frac{\partial S_i}{\partial \alpha}$. This ensures that in passing to the quantum theory we obtain a wave packet that follows the classical trajectory by forming an appropriate superposition
\beq
\Psi(\phi,t) = \int d\! \alpha f(\alpha) e^{ S_i(\phi, t; \alpha)/\hbar}.
\eeq

This wave function satisfies the Schr\"odinger equation
\beq
H_i\left(t, -i \hbar \frac{ \partial}{\partial \phi} \right) \Psi = i \hbar \frac{\partial }{\partial t} \Psi
\eeq
as a consequence of the intrinsic Hamilton-Jacobi equation.

What if we decided to employ the analogue Wheeler-DeWitt equation, using the transformed constraint,  i.e.
\beq
\frac{3}{4}   p^2_T T^2 - \frac{p_\phi^2}{2 \kappa} = 0 \rightarrow \left(- \frac{3}{4} t^2  \frac{\partial^2}{\partial t^2}  + \frac{1}{2 \kappa} \frac{\partial^2}{\partial \phi^2}\right) \Psi(\phi,t) = 0.
\eeq
rather than the intrinsic Schr\"odinger equation?

Then it turns out that if solutions are assumed of the form $\Psi = e^{i S/\hbar}$, then one can show after considerable labor, after expanding $S$ in powers of $\kappa$, that $S$ satisfies the intrinsic Hamilton-Jacobi equation. The lesson to be drawn is that the Schr\"odinger equation is far more efficient.

\subsection{Intrinsic canonical coordinate transformations}

Let us consider a point transformations of the form $T' = f(T)$. Following the general prescription described above, we set
\beq
T d\pi = - \pi' dT' + dF,
\eeq
resulting in
\beq
F = f^{-1}(T') \pi,
\eeq
and the transformed intrinsic Hamiltonian
\beq
H'_{i} =  H_{i} \frac{\partial f^{-1}(T')}{\partial T'}.
\eeq 
It must be stressed that the $T$ we have chosen in this example is a spacetime scalar only under first order four-dimensional diffeomorphisms away from the spatially isotropic solutions of Einstein's equations. Higher order variations would require the use of $R^1$ in (\ref{r1}) in constructing a spacetime scalar intrinsic time. 

The following canonical point transformation transformation transforms from York time to proper time:
$$
T' = - T^{-1} = f(T).
$$
Therefore according to our general prescription we can perform a point canonical transformation to obtain the corresponding new intrinsic Hamiltonian,
$$
p'_{T'} =  p_T \frac{d f^{-1}(T')}{dT'} = \frac{p_\phi}{\left(3 \kappa/2\right)^{1/2}} T' \frac{d (-T')^{-1}}{dT'} = \frac{p_\phi}{\left(3 \kappa/2\right)^{1/2}} \frac{1}{T'}.
$$
Note that this time  $T'$ ranges from $0$ to $\infty$.

But we need not stop here. We can actually imitate any choice of time coordinate by choosing new intrinsic times as $\lambda(T')$, for arbitrary positive definite functions $\lambda$.

This is actually a substantiation of the general result described earlier. For every choice of coordinate parameters in general relativity there corresponds a choice of intrinsic coordinates.

\subsection{Explicit invariants}
The variables that appear in the intrinsic HJ approach are precisely the invariant variables, which are constructed as follows as power series in the intrinsic time in the chosen gauge. We will display the invariant nature of the coefficients in these series.  The algebra is considerably simplified using the variable $\alpha$ defined such that $a = e^{\alpha/\sqrt{6}}$. The canonical Hamiltonian in terms of this new variable is
\beq
H_c = N e^{-3 \alpha/\sqrt{6}} \left(- \frac{\kappa}{2} p_\alpha^2 + \frac{1}{2} p_\phi^2 \right).
\eeq
The intrinsic time $T$ will be proper time if the requirement that $\dot T = 1$ fixes $N$ to be 1. We will let $T$ range from -1 to $\infty$ so that the general solution for $\phi$ will be analytic at $T = 0$. These requirements result in 
\beq
T = -\frac{1}{\kappa z} e^{z  \alpha} p_\alpha^{-1} - 1,
\eeq
where $z := 3/\sqrt{6}$. 

Following the general scheme outlined above, and presented in detail in \cite{Pons:2009ab}, the invariant observable associated with $\phi$ is 
\bea
{\cal O}_\phi &=& \phi +\left(t +h \right) \left\{ \phi, H \right\} + \left(t +h \right)^2\frac{1}{2} \left\{\left\{ \phi, H \right\},  H \right\}  \nonumber \\
&+& \left(t +h\right)^3 \frac{1}{3!}\left\{ \left\{\left\{ \phi,  H \right\},  H \right\},  H \right\} + ...  
\eea
where $h:= \frac{1}{\kappa z} e^{z \alpha} p_\alpha^{-1} + 1$.
The infinite series for the coefficient of each power of $t$ can be summed,  and the result is
\beq
{\cal O}_\phi  = \phi_0 - \frac{p_\phi}{p_\alpha \kappa z} \left(t - \frac{t^2}{2} + \frac{t^3}{3} - ...   \right) = \phi_0 - \frac{p_\phi}{p_\alpha \kappa z} \ln (1+t),
\eeq
where
$$
\phi_0(\phi, \alpha, p_\phi, p_\alpha) =  \phi -  \frac{ p_\phi}{ \kappa p_\alpha} \alpha + \frac{ p_\phi}{z \kappa p_\alpha}  \ln \left[ -z \kappa p_\alpha  \right].
$$ 
The momenta are constants, as is $\phi_0$ since in an arbitrary parameterization
\beq
\frac{d}{dt} \left( \phi -  \frac{ p_\phi}{ \kappa p_\alpha} \alpha \right) = N e^{-z \alpha} p_\phi -  \frac{ p_\phi}{ \kappa p_\alpha} \kappa N e^{-z \alpha} p_\alpha = 0.
\eeq

It is now a simple matter to find the corresponding invariant ${\cal O}'_\phi$ for other gauge choices $T' = f(T)$:
\beq
{\cal O}'_\phi = \phi_0(\phi, \alpha, p_\phi, p_\alpha) - \frac{p_\phi}{p_\alpha \kappa z} \ln \left(1+f^{-1}(t)\right).
\eeq

 Equivalence classes are fixed in this model by the value of $p_\phi$. And different choices of intrinsic time lead, as we have just observed, to different evolution.

\section{Conclusions}

 In this work we investigated the fate of background independence in a Hamilton-Jacobi approach to general relativity. A precondition for freedom from arbitrary presupposed structure must be the preservation of full diffeomorphism symmetry.
 Whereas general covariance is a basic underlying theme in configuration-velocity space, its role in phase space is still obscure. In many cases this obscurity arose with the abandonment of lapse and shift as canonical variables. As already shown in the (generally not adequately appreciated) work of  Bergmann and Komar \cite{Bergmann:1972aa}, these variables are needed in realizing the diffeomorphism group in phase space. They also necessarily appear in the construction of spacetime scalars in terms of configuration-velocity variables. Nevertheless, as shown originally by Bergmann and Komar, the explicit dependence on lapse and shift disappears when the spacetime scalars are expressed as phase space functionals.  As soon as one has in one's possession an acceptable intrinsic coordinate chart, one is then free to carry out the full diffeomorphism group of intrinsic coordinate transformations. Indeed they are canonical phase space transformations. This was Kuchar's starting point in implementing his reduced Hamiltonian idea. We have split his procedure into two steps. The first is the selection of appropriately behaved spacetime scalar functionals of the three-metric and its canonical momenta. We emphasize that this choice of new canonical configuration variables and their conjugate momenta leaves the resultant theory still fully covariant under the action of the diffeomorphism-induced canonical transformation group. And for each choice there are corresponding Einstein-Hamilton-Jacobi and Wheeler DeWitt equations. The second step is the choice of these configuration functionals as intrinsic coordinates. This is equivalent to imposing gauge conditions. And as was shown in \cite{Pons:2009ab}, the generator of diffeomorphism-induced canonical transformations can then be deployed to construct corresponding diffeomorphism invariant variables. The message we wish to communicate here is that our synthetic approach  results in the seemingly paradoxical conclusion that a choice of intrinsic coordinates yields a dynamics of diffeomorphism invariants, yet the full diffeomorphism freedom still exists since one can arbitrarily canonically transform from one intrinsic coordinate choice to another.

Thus the full four-dimensional diffeomorphism-induced group is realizeable in phase space as a canonical transformation group. Spacetime curvature-based phase space variables can be found that can serve as intrinsic  spacetime event  landmarks. The freedom in selecting these spacetime landmarks corresponds to the original diffeomorphism freedom. The dynamics expressed in terms of the intrinsic scalars, their conjugate momenta, and the remaining independent phase space variables is fully covariant under the 4-D diffeomorphism-induced group. For each choice of new phase space variables there exists a corresponding Einstein-Hamilton-Jacobi equation
$$
{\cal H}_0\left[ g_A, X^\mu, \frac{\delta S}{\delta g_B},  \frac{\delta S}{\delta X^\nu}\right] = 0.
$$
In fact, since the intrinsic coordinates must be spacetime scalars, the conventional Einstein-Hamilton-Jacobi equation will generally not be suitable for making the semi-classical transition to the Wheeler-DeWitt quantum gravitational wave equation because spacetime scalars cannot be constructed with the three-metric alone. Therefore one cannot expect a ``natural" choice of time to emerge from this equation as it did in the free particle model. Rather, the formalism must at least in a semi-classical regime admit the same full range of diffeomorphism freedom of choice of both intrinsic time and space that we find in the classical domain.  Each choice of intrinsic coordinate yields distinct explicitly temporal and spatially dependent classical equations of motion. The diffeomorphism symmetry has been employed to construct diffeomorphism invariants, and in this sense the full diffeomorphism symmetry of Einstein's theory has been taken into account. But the resulting dynamics of these invariant variables is distinct for each choice of intrinsic coordinates. In this sense the original diffeomorphism symmetry has been broken.  The quantum mechanical challenge is to construct a theory in which all of these in general unitarily inequivalent diffeomorphism invariant evolutions are taken into account. A full description of reality appears to require the collective use of all possible intrinsic times.

\begin{acknowledgements}
Thanks to Brian Pitts for his critical reading of an earlier draft of this paper. D. S. would also like to thank Nicholas Wheeler for inspiring his interest in Hamilton-Jacobi theory.
\end{acknowledgements}

\section*{Appendix A: The free particle principal function}

The true Hamilton-Jacobi equation for the free relativistic particle is obtained from the increment (\ref{dSpart}). In this case we know the general solution, 
\beq
q^\mu(\theta) = \bar q^\mu + \frac{\bar p^\mu}{m} f(\theta),
\eeq
where the barred quantities are constants,  and $f(\theta)$ is monotonically increasing but otherwise an arbitrary function.  Given the general solution the corresponding increment in the action can be obtained by merely substituting into (\ref{dSpart}), obtaining
\beq
S(q, \theta; \bar p) = \bar p_\mu q^\mu - \frac{1}{2} (\bar p^2 + m^2) f(\theta) - C(\bar p).
\eeq
We confirm that the Hamilton-Jacobi equation (\ref{HJpart}) is satisfied. Furthermore, setting 
$$
0 = \frac{\partial S}{\partial \bar p_\mu} = q^\mu - f(\theta) \bar p^\mu - \frac{\partial C}{\partial \bar p_\mu},
$$ 
we recover the exact general solution -  but the constraint $\bar p^2 + m^2 = 0$ must be applied \underline{after} differentiation with respect to $\bar p_\mu$,  and on this surface in phase space the $\theta$ dependence disappears.

\section*{Appendix B: Critique of the Dittrich multi-fingered time approach to Dirac observables}

Dittrich has pioneered a technique for constructing in generally covariant theories objects which are invariant under the action of the secondary first class constraints. We learned from her how to Abelianize these constraints so as to be able to easily find the phase space dependent coordinate transformations that yield these invariants. But we approach the problem from a distinctly different conceptual perspective. The essential difference is that she employs passive coordinate transformations to construct her invariants whereas we employ active canonically realized diffeomorphism-induced symmetry transformations. She does not take this route since she apparently does not recognize that the full four-dimensional symmetry group is of relevance in this context. Indeed, in order to be able to implement this group the lapse and shift functions must be retained as canonical phase space variables. Connected with this observation is the fact that the Hamiltonian that generates evolution in time is not to be confused with the generator of diffeomorphism-induced symmetry transformations. As must be demanded of a symmetry, the latter actually transforms solutions of the equations of motion into new solutions. On the other hand, in pursuing the Kucha\v{r} multi-fingered time approach Dittrich confines her attention to the Hamiltonian - and thus does not reproduce this fundamental symmetry requirement in her canonical formalism.

The evolution in the Dittrich-Kucha\v{r} multi-fingered time approach  is understood in general relativity as a ``pushing forward in time'' described by an embedding mapping from $\Sigma \times \Re \rightarrow {\cal M}$, where $\Sigma$ is a three dimensional manifold. Using adapted coordinates, the mapping is $Z^\mu_t(\vec x) = x^\mu$. The basic assumption is that one decompose the ``push forward''  in terms of lapse and shift as
$$
\delta x^\mu = \left(N n^\mu  + \delta^\mu_a N^a \right) \delta t,
$$
where the normal to the hypersurface is $n^\mu = N^{-1} \delta^\mu_0 - \delta^\mu_a N^{-1} N^a$. This decomposition of the push forward is simply assumed in the Kucha\v{r} formalism. We now know that it is required in order to be able to project configuration-velocity transformations under the Legendre map to phase space \cite{Pons:1997aa}.
The crucial difference with the fully diffeomorphism covariant Hamiltonian approach that we describe in this paper is that any changes in lapse and shift must be inserted in the multi-fingered time approach  ``by hand''; they are not canonically generated and in this sense the full four-dimension diffeomorphism covariance is lost.

We will detail these differences using as a simple example the free relativistic particle model. For this purpose, since it corresponds more closely to general relativity,  it will be instructive to formulate the model using an auxiliary variable $N$ that actually serves as a lapse function on the parameter space. Thus we take the Lagrangian to be
$$
L = \frac{1}{2N} \dot q^2 - \frac{1}{2} m^2 N.
$$
Then we have the primary constraint $\pi = 0$, where $\pi$ is the momentum conjugate to $N$, and the secondary constraint $C_\bot =\frac{1}{2} \left( p^2 + m^2\right) = 0$. The Dittrich Hamiltonian would be 
$$
H[N]  =  N C_\bot.
$$
(The embedding looses dependence on the lapse in this one-dimensional case.)
Then given $N(\theta)$ one solves the Hamiltonian equations
$$
\frac{dq^\mu}{d\theta} = N(\theta) p^\mu,
$$
and $\frac{dp^\mu}{ds} = 0$. 
The crucial observation here is that $N(\theta)$ does change its form under reparameterizations, but these changes are not generated through canonical transformations. There does however exist a means of performing a passive parameter transformation that will alter $N$. Suppose that $N(\theta)$ is given, then the corresponding particle solution is
$$
q^\mu(\theta) = q^\mu(\theta_0) +p^\mu \int_{\theta_0}^\theta d\theta' N(\theta').
$$
One can then carry out the Dittrich complete observable program by selecting, for example, to put  the partial variable $q^0(\theta)$ in correspondence with the partial variables $q^a(\theta)$. This correspondence qualifies as an intrinsic parameter choice since $q^0$ is a reparamterization scalar. Thus we can conceive of $\hat \theta= f(\theta):= q^0(\theta) = q^0(\theta_0) +p^0 \int_{\theta_0}^\theta d\theta' N(\theta')$ as a reparameterization. Then since $N$ transforms as a scalar density of weight minus one, we can calculate the passively transformed $N$ through the formula
$$
\hat N (\hat \theta) = N(\theta) \frac{d \theta}{d \hat \theta}.
$$
The actively transformed $N$ is then
$$
\hat N ( \theta) = \frac{N \left( f^{-1} (\theta) \right)}{\left. \frac{ d f (\theta)}{d \theta} \right|_{f^{-1}(\theta)}}.
$$
Dittrich can carry out the passive transformation, but she is not able to produce the active result through a canonical transformation since she does not employ the full parameterization-induced generator. This generator is
$$
G(\xi) = \xi H + \dot \xi \pi.
$$
It generates infinitesimal canonical transformations that follow from infinitesimal reparameterizations of the form $\hat \theta =\theta + \delta \theta = \theta - \frac{\xi(\theta)}{N(\theta)}$. With this generator one can with a finite $\xi(\theta)$ generate a one-parameter family of  phase space solutions $q_s^\mu(\theta)$, $N_s(\theta)$ from any given initial set of solutions. In other words, this generator exploits the underlying Lagrangian reparameterization symmetry and does indeed transform solutions of the Euler-Lagrange equations into new solutions. This is an essential feature that is lacking in Kucha\v{r} multi-fingered time approach in which one deals only with the lapse and shift in advancing the evolution time. Only in the case of three-dimensional diffeomorphisms is the symmetry fully implemented in the Kucha\v{r} embedding procedure as a canonical transformation.

Let us also point out that since we can implement reparameterizations, we can now check explicitly whether reparameterization scalars transform correctly. We find for example that
$$
\bar \delta q^\mu = \left\{q^\mu, G(\xi)  \right\} = p^\mu \xi = \dot q^\mu \frac{\xi}{N} = -\dot q^\mu \delta \theta.
$$
We contrast this demand with the Kucha\v{r} requirement that a variable transform as a scalar under reparameterizations (or in general relativity as a spacetime scalar). Since the timelike transformations are not implemented as a symmetry, Kucha\v{r} was forced to characterize a spacetime scalar $S(\vec x)$ as a variable whose variation satisfied  $\left\{S(\vec x) , \int d^3 x C_\bot (\vec x) N (\vec x) \right\} = 0$ when $N (\vec x') = 0$ \cite{Kuchar:1976aa,Kuchar:1976ab,Dittrich:2006aa} Actually this is a necessary condition, but it is not sufficient. The variable $v_\mu n^\mu$, for example, that Bergmann has termed a ``D-invariant'' satisfies this requirement, and it is is clearly not a spacetime scalar \cite{Bergmann:1962ac}.

One might wonder how the Hamiltonian transforms under reparameterizations. The Dirac Hamiltonian is
$$
H_D = N H + \lambda \pi,
$$
where $\lambda $ is a positive-definite but otherwise arbitrary function of $\theta$. Since it does not depend on the phase space variables one might wonder how it transforms under the action of $G(\xi) $. It does so because it is explicitly $\theta$-dependent. The transformed Hamiltonian is
$$
\hat H = H + \{H, G(\xi) \} + \frac{\partial G}{\partial \theta} =H+  \dot \xi H + \ddot \xi \pi.
$$
But under this transformation $\hat N = N + \dot \xi$, $\hat q^\mu = q^\mu + \xi p^\mu$ while $p^\mu$ and $\pi$ are unchanged. Substituting we find
$$
\hat H_D = \hat N \hat H + \left( \lambda + \ddot \xi \right) \hat \pi.
$$
Thus $\lambda$ has varied correctly so that $\frac{d \hat N}{d \theta} = \frac{d N}{d \theta} + \frac{d^2 \xi}{d \theta^2}$. 

These results can be easily generalized to any generally covariant theory, including general relativity.

\bibliographystyle{plain}
\bibliography{qgrav-V19}

%
%

\end{document}